# Strong Hyperfine-Induced Modulation of an Optically-Driven Hole Spin in an InAs Quantum Dot

Samuel G. Carter,[1] Sophia E. Economou,[1] Alex Greilich,[2][†] Edwin Barnes,[3] Timothy M. Sweeney,[4] Allan S. Bracker,[1] and Daniel Gammon[1]

[1] *Naval Research Laboratory, Washington, DC 20375, USA*

[2] *Contractor, University of Maryland, College Park, MD 20742, USA; resided at the Naval Research Laboratory, Washington, DC 20375, USA*

[3] *Condensed Matter Theory Center, Department of Physics, University of Maryland, College Park, MD 20742-4111, USA*

[4] *NRC postdoctoral associate residing at the Naval Research Laboratory, Washington, DC, 20375, USA*

[†] *Present address: Technische Universität Dortmund, 44221 Dortmund, Germany*

Compared to electrons, holes in InAs quantum dots have a significantly weaker hyperfine interaction that leads to less dephasing from nuclear spins. Thus many recent studies have suggested that nuclear spins are unimportant for hole spin dynamics compared to electric field fluctuations. We show that the hole hyperfine interaction can have a strong effect on hole spin coherence measurements through a nuclear feedback effect. The nuclear polarization is generated through a unique process that is dependent on the anisotropy of the hole hyperfine interaction and the coherent precession of nuclear spins, giving rise to strong modulation at the nuclear precession frequency.

The hyperfine interaction between electron spins and nuclear spins in solid state materials has been of particular interest over the past decade, largely driven by a desire to take advantage of electron spins as quantum bits [1]. For electron spins in quantum dots (QDs), the hyperfine interaction with a bath of nuclear spins is the dominant source of dephasing as fluctuations in the nuclear polarization induce fluctuations in the electron spin precession frequency through the Overhauser shift [2,3]. For holes, however, the contact hyperfine interaction, which dominates for electron spins, is suppressed due to the *p* orbital symmetry of the top of the valence band. The dipole-dipole hyperfine interaction is still present, but it is an order of magnitude weaker and is very anisotropic [4–9], leading to hopes of a highly coherent hole spin qubit.

Recent experimental results in InAs QDs have yielded several disparate values for the heavy hole inhomogeneous dephasing time $T_2^*$, ranging from 2 ns to values exceeding 100 ns [10–15], significantly longer than the 1-2 ns dephasing time for electron spins in these QDs [16]. The homogeneous dephasing time $T_2$ has been measured as ~1 μs [13,17], with a spin relaxation time of ~1 ms [18,19]. Experimental techniques and capabilities developed for electron spins in QDs, including ultrafast optical single qubit and two qubit gates in coupled QDs [20], have also been demonstrated for hole spins [12]. While the hole spin shows promise as a spin qubit [21–23], the influence of nuclear spins on hole spin dynamics and the ability of holes to generate a nuclear polarization are still unclear. Because the hole hyperfine interaction is weaker compared to electrons, much recent work has suggested that nuclear spins are unimportant in hole spin dynamics compared to the effects of fluctuating electric fields.

In this Letter we demonstrate that nuclear spins can have a strong effect on optically-controlled hole spins in QDs. We present Ramsey fringe and spin echo measurements of a single heavy hole spin in an InAs QD that show a strong amplitude modulation at the nuclear spin precession frequency. The modulation frequency scales with the external magnetic field and matches twice the Larmor precession frequency of indium nuclear spins. Contrary to intuition, the effect is not due to a time-dependent Overhauser shift of the hole spin at the nuclear precession frequency since the hole hyperfine interaction is too weak. Instead, the hole spins generate a significant nuclear spin polarization through a novel mechanism that depends on the phase of nuclear spins precessing in an external magnetic field and the anisotropy of the interaction. This nuclear polarization then indirectly affects the hole spin by shifting the optical transitions of the hole spin system through the Overhauser shift of the *electron* in the positively charged trion, which is much greater than that of the hole. A theoretical model provides qualitative agreement with the experimental results. These results demonstrate that nuclear spin effects can be quite important in hole spin dynamics and that coherent precession of nuclear spins can strongly influence the polarization process.

Our experiments are performed on a single pair of vertically-stacked, coupled InAs QDs that are embedded in a p-type Schottky diode (see Supplemental Material or Ref. [12] for further details). The sample is held at a temperature of 6-7 K with the magnetic field $B$ perpendicular to the growth direction. In the experiments reported in this work, the bias is such that the bottom QD is charged with two holes and the top QD is charged with a single hole. With the two bottom QD holes in the singlet state and minimal coupling between QDs at this bias, the only spin degree of freedom is for the single heavy hole spin in the top QD. We therefore consider the system as a single QD charged with a single hole. The presence of the bottom QD has the advantage that the optical linewidths (8 µeV) are significantly improved over our single QD samples, which we tentatively attribute to screening of charge fluctuations of nearby beryllium acceptors [24,25].

The hole spin is initialized, controlled, and measured using the optical transitions from the two hole spin states to the two charged exciton (trion) states displayed in the inset of Fig. 1(a). We make use of the Lambda system formed by the two spin states of the hole and the upper trion state. Hole spin dynamics are first measured using Ramsey interference fringes, in which the delay between two optical spin rotation pulses is varied while measuring the resulting spin population [26]. In the absence of nuclear feedback, the result should give exponentially decaying sinusoidal oscillations at the Larmor precession frequency of the hole.

The pulse sequence for the Ramsey interference fringes is displayed in Fig. 1(a). First, a 20 ns narrowband initialization pulse from a cw tunable diode laser modulated by an acousto-optic modulator (AOM) is tuned to transition 1, which optically pumps the QD into the hole spin up state $|\Uparrow\rangle$ [19,27,28]. After 30-50 ns delay, a short (13 ps, 150 µeV bandwidth), circularly-polarized pulse from a mode-locked laser is used to optically rotate the spin state by $\pi/2$ about the optical axis. The pulse is detuned below the lowest energy transition by 200-300 µeV to rotate the hole spin state through a virtual process [26,29–34]. A second short pulse, with a variable delay $\tau$ from the first, again rotates the hole spin by $\pi/2$ about the optical axis, which brings the hole spin closer to $|\Uparrow\rangle$ or $|\Downarrow\rangle$ on the Bloch sphere, depending on the spin phase at the second pulse. Another narrowband 20 ns pulse measures the differential transmission (DT) [35] of transition 1 to determine the spin population and to reinitialize the system as part of the next pulse sequence. The pulse sequence period $T_{\text{cycle}}$ is 123 ns.

Figure 1(b) displays Ramsey fringes at a magnetic field of 4 T, showing hole spin precession at 4 GHz (g-factor of 0.071) and a slow amplitude modulation at ~70 MHz. The inset of Fig. 1(b) plots these same Ramsey fringes over a small delay range from 2-3 ns to better display hole spin fringes. The fringes have a sawtooth shape that we attribute to a nuclear polarization that depends on the pulse delay. The nuclear polarization shifts the optical transitions relative to the laser, primarily due to the Overhauser shift of the unpaired electron spin in the trion state, which is significantly larger than that of the hole spin states. These shifts weaken the initialization and measurement process, and distort the fringes. This delay-dependent nuclear polarization effect is similar to that observed for electron spins [36] and will be used to explain the slow modulation.

A striking feature of our data is the strong amplitude modulation of the fringes. Figure 1(c) plots the fringe amplitude vs. delay, obtained by fitting every two fringes to a cosine, showing a strong modulation at 70 MHz with a decay time $T_2^*$ of 19 ns. We attribute the modulation to a significant nuclear polarization that is generated when the delay $\tau$ matches half-integer multiples of the nuclear precession period (at 0, 12, and 26 ns), as will be explained below. With several different nuclear species present ($^{69}$Ga, $^{71}$Ga, $^{75}$As, $^{113}$In, and $^{115}$In), modulation could occur at several frequencies, but $^{115}$In is expected to dominate due to its large nuclear spin of 9/2 and high abundance. The expected $^{115}$In precession frequency is 37.5 MHz at 4 T [37], which is roughly half the observed modulation frequency. In Fig. 1(d), a theoretical calculation of the Ramsey fringes which takes into account the interaction with nuclear spins qualitatively reproduces this modulation, as will be discussed below. The modulation is also observed at a magnetic field of 6 T with a frequency of 112 MHz, roughly scaling with magnetic field.

Spin echo measurements were performed to further characterize this behavior and eliminate the effects of inhomogeneous dephasing. As displayed in Fig. 2(a), a higher intensity pulse yielding a $\pi$-rotation of the hole spin is inserted between the two $\pi/2$ pulses, with nearly equal delays of $T-\delta T$ and $T+\delta T$ between adjacent pulses. We expect inhomogeneous dephasing of this system due to fluctuations in the nuclear spin polarization or in the charge environment. If these fluctuations are slow compared to $T$, the $\pi$-pulse rephases the system, giving a spin echo. The echo amplitude is measured by scanning a small delay offset $\delta T$ that shifts the echo relative to the second $\pi/2$ pulse by $2\delta T$ [16]. Typically, the spin echo will monotonically decay as a function of $T$, according to the homogeneous dephasing time $T_2$. Figure 2(b) plots the spin echo signal at 4 T as a function of $\delta T$ for a series of echo pulse delays $T$. The amplitude and phase of the echo are significantly modulated as a function of $T$. The amplitude of the echo is plotted as a function of $T$ in Fig. 2(c), with a fit to a decaying sinusoidal function, $A\exp(-2T/T_2)(1-\cos\Omega T)$, that gives $T_2 = 74$ ns and $\Omega = 73.6$ MHz. This modulation has nearly the same frequency and presumably the same origin as the modulation in the Ramsey fringes.

Spin echo measurements were also performed for a series of magnetic fields from 3-6 T to test how well the modulation scales with magnetic field. The amplitudes vs. $T$ for all fields are plotted in Fig. 2(d), with the delays scaled to $B = 4$ T for each data set. (The delays at magnetic field $B$ were scaled $T \rightarrow (B/4)T$.) Most of the points fall on the decaying oscillation rather well with fit parameters of $T_2 = 58$ ns and $\Omega = 74.6$ MHz, indicating that the modulation scales with magnetic field as expected for a modulation mechanism that depends on nuclear spin precession.

At first glance, the Ramsey fringe and echo modulation observed here might naively be attributed to a precessing nuclear spin polarization that periodically modifies the hole spin precession frequency, which is commonly observed for electron spins in solid state systems [38]. For this effect to occur, a significant hyperfine interaction is necessary as well as nuclear precession about a different quantization axis than that of the resident electrons or holes. This type of spin echo modulation has been observed for electrons in electrically-defined QDs in GaAs at low magnetic fields, disappearing above 0.2 T [39], and theoretically predicted for holes in QDs at low magnetic fields [40]. These effects become negligible at the high magnetic fields used here, where the hole and nuclear spin quantization axes are fixed by the strong field and the hole hyperfine interaction is weak relative to the Zeeman terms.

Instead, we attribute the modulation observed here to the generation of a significant nuclear polarization when the optical pulse delay $\tau$ matches the precession of indium nuclear spins. As illustrated in Fig. 3(a), the nuclear polarization shifts the optical transition out of resonance with the initialization/measurement laser due to the Overhauser shift of the trion spin states. This periodically reduces the generated hole spin polarization and thus the amplitude of the fringes. For a nuclear polarization oscillating positive and negative each period, the amplitude is reduced twice per period. (Note that the nuclear polarization is not changing on a nanosecond timescale. Instead, a different polarization builds up at each pulse delay, which is changed about once per second.)

The origin of the nuclear polarization can be understood conceptually from a physical picture with two spins, one hole and one nuclear spin, with a Hamiltonian $H = \omega_h S_z + \omega_N I_z + A_x^h S_x I_x$. The hole and nuclear spin splittings are $\omega_h$ and $\omega_N$ with spin operators $S$ and $I$, and the hyperfine interaction $A^h$ is approximated as Ising-like. The effects of optical pulses are not included in this picture, but they act to collapse the system after the evolution time. The evolution can be described by two pseudospins, one corresponding to the two states in which the hole and nuclear spins are parallel and one corresponding to the two states in which they are antiparallel. The Bloch spheres for these pseudospins are shown in Fig. 4. The two Bloch vectors evolve independently in the absence of optical pulses and with precession frequencies $\omega_h \pm \omega_N$. The hyperfine interaction causes the Bloch vectors to precess about an axis tilted with respect to the North-South axis. If we start in $|\Uparrow\downarrow\rangle$ the pseudospin precesses about the tilted axis, giving a small time-dependent probability of $|\Downarrow\uparrow\rangle$. Measurement of the hole spin projects the system into $|\Uparrow\downarrow\rangle$ or $|\Downarrow\uparrow\rangle$, giving a small probability of a hole-nuclear flip-flop $w_+$ that oscillates at $\omega_h - \omega_N$. The same occurs starting in $|\Uparrow\uparrow\rangle$, with a flip flop rate $w_-$ that oscillates at $\omega_h + \omega_N$. The difference in these spin flip rates oscillates at $\omega_N$, which gives rise to a nuclear polarization that depends on the hole-nuclear evolution time.

While the main qualitative feature of the data, the modulation at the indium precession frequency, can be understood via this physical picture, a more sophisticated approach along the lines of Ref. [41] is needed to model the full dynamics of the hole as modified by the net nuclear bath. The main idea of the theory is that the laser-generated hole polarization is transferred to the nuclear spins via the joint hole-nuclear evolution under the hyperfine interaction and the B-field. Subsequent hole spin measurement disentangles the joint quantum state and collapses the nuclear polarization. At each delay $\tau$, a steady state nuclear polarization $P = N\langle I_z \rangle$ is generated over many pulse repetition cycles, where $N$ is the number of indium nuclei. The nuclear polarization relies on optical excitation of the trion, so it depends on the detuning $\Delta$ of the narrowband laser from the QD transition. But this detuning itself depends on the nuclear polarization through the Overhauser shift, giving rise to a feedback effect. $\Delta$ and $P$ must be found self-consistently, solving $\Delta(\tau) = A_z^e P(\Delta, \tau)$, where $A^e$ is the electron hyperfine interaction. For a full mathematical description of the theory see the Supplemental Material.

In Fig. 1(d) we plot the spin component of the hole along the B-field as a function of the delay, simulating the Ramsey fringes. Figure 3(b) displays the Overhauser shift of the trion, showing the fast oscillations that give the sawtooth shape and the slow oscillations that give the modulation at the nuclear frequency. We can thus reproduce the salient features of the experiment. The theory also shows that an isotropic hyperfine interaction gives essentially no fringe modulation, and larger hole g-factors reduce the modulation. Since these parameters are quite dependent on the geometry of the QD, this may explain why a similar experiment with different QDs by another group did not observe this effect [13]. Modulation of the echo signal when a π-pulse is added also occurs in our theory with the same physical explanation. However, the change in phase with $T$ observed in Fig. 2(b) is not produced by the theory and is not yet understood.

The results presented in this Letter demonstrate that while the hole hyperfine interaction is weaker than for electrons, resulting in a longer $T_2^*$, the hole hyperfine interaction can still generate a nuclear polarization that has a strong effect on hole spin measurement and initialization. The generation of the nuclear polarization is a result of the anisotropic hyperfine interaction that is stronger along the growth direction. The strength of the nuclear polarization is also sensitive to the phase of the nuclear spin precession with respect to the optical pulses, which, to our knowledge, is the first observation of such a phenomenon. Interestingly, our theoretical model indicates that the direct effect of this nuclear polarization on the hole spin is minimal, and that the dominant effect on the hole is from the Overhauser shift of the electron spin in the trion state, which shifts the optical transition frequency out of resonance with the initialization and measurement laser and thus reduces the hole spin amplitude. This suggests that the $T_2^*$ measured here is not limited by nuclear spin fluctuations but instead by charge fluctuations, as discussed previously [12]. However, the measured echo decay time $T_2$ is shorter than values obtained in a similar system [13], which may be due to inhomogeneity of the indium nuclear precession frequency and/or contributions from gallium and arsenic nuclear spins that wash out the echo. The improved understanding of this rich physical system developed here should significantly help in improving hole spins as qubits and could also be used to optically control the nuclear spin polarization [36,42,43].

This work was supported by a Multi-University Research Initiative (US Army Research Office; W911NF0910406), the NSA/LPS, and the US Office of Naval Research.

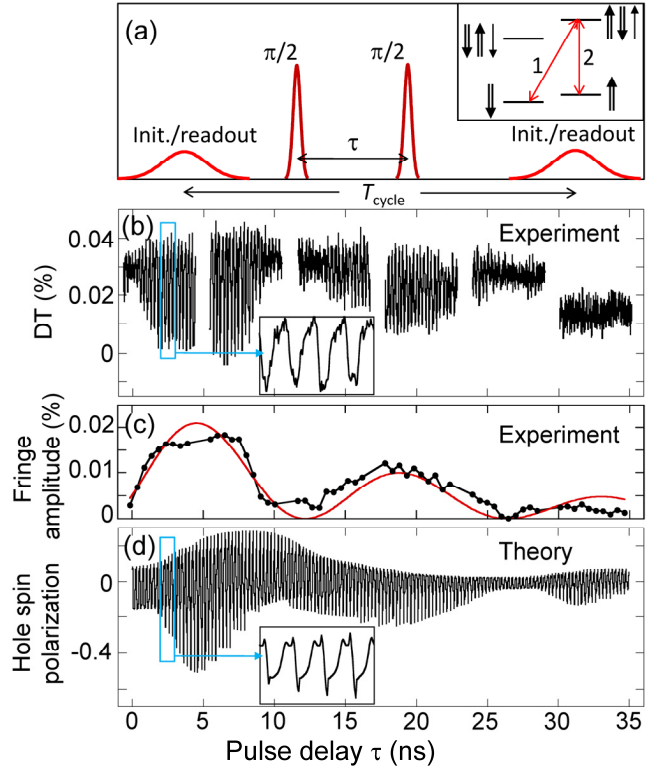

Fig. 1 (color online). (a) Pulse sequence for Ramsey interference fringe measurements. The inset displays the hole-trion level diagram in a Voigt magnetic field, with single (double) arrows representing electron (hole) spins. (b) Ramsey fringes at 4 T. (c) Fringe amplitude vs. delay obtained from the data in (b), with a decaying cosine fit. (d) Theoretically calculated Ramsey fringes at 4 T multiplied by an exponential decay with a 20 ns decay time. The hole hyperfine interaction is nearly Ising along the growth direction ($A_x^h = 1.05$ neV), with an anisotropy of 90%. The insets in (b) and (d) magnify a 1 ns delay window for experiment and theory to display the fringe shape.

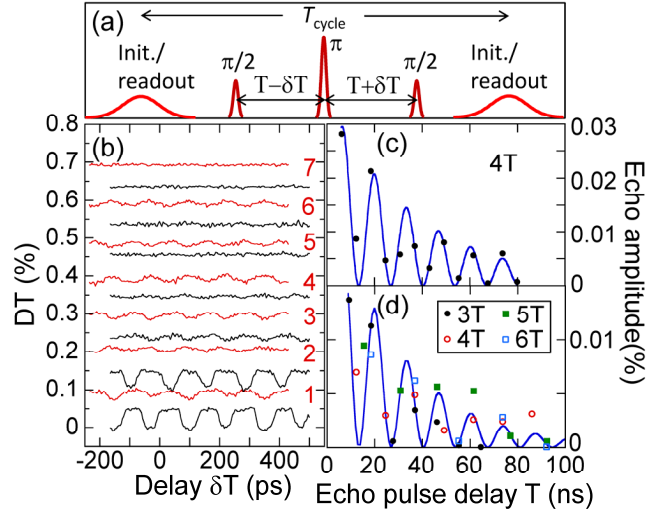

Fig. 2 (color online). (a) Pulse sequence for hole spin echo measurements. (b) Ramsey fringe echo signal at B = 4 T as a function of δT for a series of values of $T$, which are labeled in units of $T_R = 12.3$ ns for red/gray curves. Black curves are half-integer steps of $T_R$. $T_{cycle}$ = 246 ns (c) Echo amplitude vs. $T$, obtained from fitting the data in (b). (d) Echo amplitude vs. $T$ for several magnetic fields, in which the $T$ is scaled by the magnetic field according to $T \to (B/4)T$. Fits of the data in (c) and (d) to a decaying cosine are shown as solid lines.

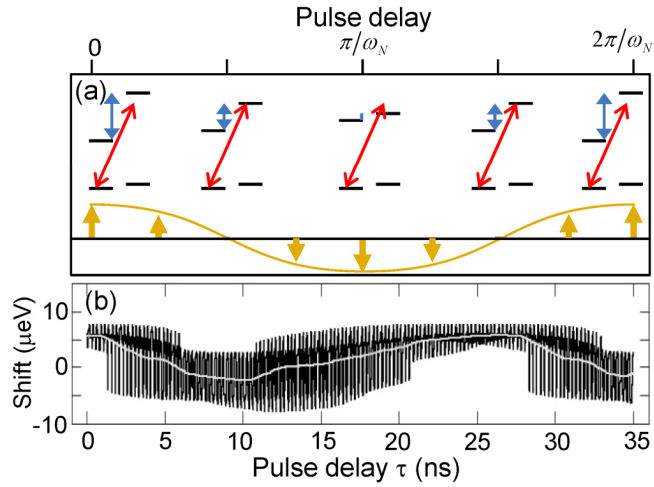

Fig. 3 (color online). (a) Illustration of how the nuclear polarization (yellow arrows) modulates initialization and measurement of the hole spin through the Overhauser shift of the trion. The level diagrams display the hole (lower) and trion (upper) energy levels and the initialization/readout laser for different nuclear polarizations. (b) Theoretically calculated Overhauser shift of the trion for the same parameters as in Fig. 1(d), with the white line smoothing over the fast oscillations.

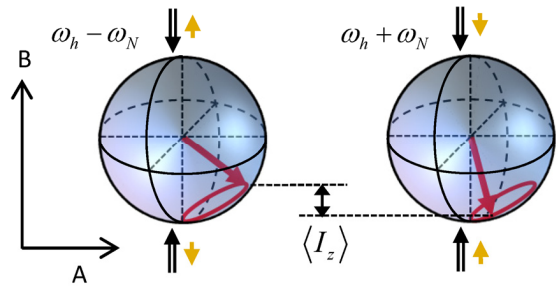

Fig. 4 (color online). The two Bloch spheres representing the interacting hole-nuclear spin system that gives rise to a nuclear polarization, with double arrows representing the hole spin and short single arrows representing a nuclear spin. The Ising-like hyperfine interaction $A$ is perpendicular to the quantization axis along the magnetic field $B$. The equal statistical mixture of both pseudospins pointing along the south poles represents a fully polarized hole and an unpolarized nuclear spin.